# Automatic Speech Recognition in Sanskrit: A New Speech Corpus and Modelling Insights

**Devaraja Adiga**[1*]
pdadiga@iitb.ac.in

**Rishabh Kumar**[1*]
krrishabh@cse.iitb.ac.in

**Amrith Krishna**[2]
ak2329@cam.ac.uk

**Preethi Jyothi**[1]
pjyothi@cse.iitb.ac.in

**Ganesh Ramakrishnan**[1]
ganesh@cse.iitb.ac.in

**Pawan Goyal**[3]
pawang@cse.iitkgp.ac.in

[1]IIT Bombay, Mumbai, India; [2]University of Cambridge, UK; [3]IIT Kharagpur, WB, India

## Abstract

Automatic speech recognition (ASR) in Sanskrit is interesting, owing to the various linguistic peculiarities present in the language. The Sanskrit language is lexically productive, undergoes euphonic assimilation of phones at the word boundaries and exhibits variations in spelling conventions and in pronunciations. In this work, we propose the first large scale study of automatic speech recognition (ASR) in Sanskrit, with an emphasis on the impact of unit selection in Sanskrit ASR. In this work, we release a 78 hour ASR dataset for Sanskrit, which faithfully captures several of the linguistic characteristics expressed by the language. We investigate the role of different acoustic model and language model units in ASR systems for Sanskrit. We also propose a new modelling unit, inspired by the syllable level unit selection, that captures character sequences from one vowel in the word to the next vowel. We also highlight the importance of choosing graphemic representations for Sanskrit and show the impact of this choice on word error rates (WER). Finally, we extend these insights from Sanskrit ASR for building ASR systems in two other Indic languages, Gujarati and Telugu. For both these languages, our experimental results show that the use of phonetic based graphemic representations in ASR results in performance improvements as compared to ASR systems that use native scripts.[1]

## 1 Introduction

Sanskrit is a language with fairly advanced disciplines of phonetics (*Śikṣā*), prosody (*Chandas*), and grammar (*Vyākaraṇa*). The language has a rich oral tradition and it tends to follow a phonemic-orthography resulting in systematic grapheme-phoneme correspondences. Connected speech leads to phonemic transformations in utterances, and in Sanskrit this is faithfully preserved in writing as well (Krishna et al., 2018). This is called as Sandhi and is defined as the euphonic assimilation of sounds, *i.e.*, modification and fusion of sounds, at or across the boundaries of grammatical units (Matthews, 2007, p. 353). Phonemic orthography is beneficial for a language, when it comes to designing automatic speech recognition Systems (ASR), specifically for unit selection at both the Acoustic Model (AM) and Language Model (LM) levels.

Regardless of the aforementioned commonalities preserved in both the speech and text in Sanskrit, designing a large scale ASR system raises several challenges. The Unicode encoding for the native scripts in Sanskrit, both in Roman and Devanāgari, does not preserve the correspondence with the phonemic encoding. Further, mapping the graphemes in Unicode to the corresponding phonemes either leads to ambiguity and redundancy or often requires multi-grapheme combinations.

The language is lexically productive, which results in long compound words with multiple components in usage. This results in the speakers segmenting the compounds at arbitrary lexeme boundaries of the compound, as it need not always be possible to utter the compound in one breath and also to convey the meaning clearly. Similarly, such arbitrary segmentations at the word boundaries are possible in utterance of long text sequences where multiple lexical items are fused together via Sandhi. These segmentations are accompanied with the corresponding Sandhi based transformations, resulting in a new phonetic sequence different from the original sequence. Finally, Sanskrit might be one of those rare natural languages where the number of non-native proficient speakers are manifold in comparison to the native speakers (Krishna et al., 2020). This makes the ASR task further challenging, as the speakers are prone to carry their influence from their corresponding mother tongues into the Sanskrit utterances as well.

---

[*]Joint first author
[1]Dataset and code can be accessed from www.cse.iitb.ac.in/~asr and https://github.com/cyfer0618/Vaksanca.git.

While there exist several computational models for processing Sanskrit texts (Kulkarni, 2013; Kumar et al., 2010; Shukla et al., 2010; Kulkarni et al., 2010a; Goyal et al., 2012; Kulkarni et al., 2010c; Mishra et al., 2013; Saluja et al., 2017; Anoop and Ramakrishnan, 2019; Krishna et al., 2021), large scale systems for processing of speech in Sanskrit, are almost non-existent. First, we present a new dataset, with 78 hours of speech covering about 46,000 sentences, for ASR in Sanskrit. Keeping the rich and long cultural heritage the language carries, we prepare our dataset to be diverse both chronologically and in terms of the domain coverage. Further, the dataset contains utterances from 27 different speakers, representing 6 different native languages. The dataset splits have disjoint speakers, with 12 in the training and 5 each in the validation, test and out-of-domain test data sets. Further, we explicitly mark the segmentation decisions made by a speaker to segment long compound words and fused phrases and include the corresponding transformations due to sandhi.

Using this dataset, we propose a new, large-vocabulary Sanskrit ASR system, which, to the best of our knowledge, is the first such system for Sanskrit. The phonemic orthography followed in Sanskrit has influenced our design choices in terms of unit selection at the level of the acoustic and language models. We investigate three different encoding schemes used to model LM tokens, namely, word-based encoding, byte pair encoding (BPE) and a new vowel split encoding inspired by existing linguistic theories of syllabic structure popularly used within text-to-speech systems (Kishore and Black, 2003; Mishra et al., 2013). Further, to address the redundancy issues in Unicode representations, we make use of the Sanskrit Library Phonetic (SLP1) encoding scheme proposed by Scharf and Hyman (2011). SLP1 is designed such that it preserves the phonemic orthography. Building on the study by Scharf and Hyman (2011), we focus on two graphemic representations only, viz., native script (Devanagari) and SLP1.

Finally, we extend our insights to model ASR systems for two more Indian languages, viz., Telugu and Gujarati. We extend the SLP1 to include graphemes relevant for these languages which are missing from Sanskrit. We report the performance of these ASR systems on two publicly available ASR datasets.

Our main contributions in this work are:
1) We present (in Section 2) a new, large vocabulary Sanskrit ASR system and the first ever ASR-based study for Sanskrit using a new, large and diverse, labeled speech corpus वाक्सञ्चयः (/Vāksañcayaḥ/).
2) We investigate (in Sections 3 and 4) different modeling choices for both acoustic models and language models in Sanskrit ASR systems, along with different graphemic representations. We propose a new word segmentation technique based on splitting at vowels that can be used with both the acoustic model and the language model.
3) We also contextualize our findings for Sanskrit by providing comparisons on ASR systems built for two other Indian languages, viz., Gujarati and Telugu.

## 2 A new Sanskrit Speech Corpus: वाक्सञ्चयः(/Vāksañcayaḥ/)

Our corpus वाक्सञ्चयः (/Vāksañcayaḥ/), has more than 78 hours of data with an overall vocabulary size of 91,000 words and recordings of about 46,000 sentences, each with a sampling rate of 22 KHz. The contents span over 3 time periods categorised into pre-classical literature (1,500 BCE - 100 BCE), classical literature (300 CE - 800 CE) and modern literature (900 CE to now). The corpus is intended to address the challenges in interfacing the speech and the text covered by the disciplines of phonetics (Śikṣā), and grammar (vyākaraṇa). Hence, we confine our corpora to those written only in prose (Gadya)[2]. In the Sanskrit literature, frequency of commonly used words changes from one topical domain to another, specifically one *Śāstra* (branch of learning) to another (Adiga et al., 2018). Our corpus contain samples from diverse domains, including philosophy, literature, commentary on poetry and grammar. It also includes contemporary recordings such as stories, live lectures, spiritual discourse and radio program/podcast, so that collecting a wide range of Sanskrit vocabulary.

| Dataset | Speakers | Hours | Utterances |
|---|---|---|---|
| Train | 12 | 56 | 34,309 |
| Validation | 4 | 7 | 3,190 |
| Test | 6 | 11 | 6,004 |
| Out-of-domain Test | 5 | 5 | 2,618 |

Table 1: Overview of Sanskrit speech corpus.

---

[2]We do not include verses in our current dataset, as modelling ASR systems for verses would require additional resources on both the acoustic model and the language model fronts.

The recordings were primarily collected with the help of volunteers, recording their speech by using the Recorder app on Android phones and the Audacity platform, and from various sources available online.[3] oTranscribe[3] was used to transcribe the audio files. We had a total of 9 volunteers for recording and 18 unique speakers for the content collected online. Each of these speakers are proficient Sanskrit speakers, with at least an undergraduate or equivalent degree in Sanskrit. These speakers are native speakers of one of the 6 following Indic languages, Hindi, Kannada, Malayalam, Marathi, Tamil and Telugu. In Table 1, we provide the details of the training/validation/test splits for our corpus, वाक्सञ्चयः (/Vāksañcayaḥ/). The speakers in all these 4 splits, train, validation, test, and out-of-domain test sets are disjoint. The out-of-domain test dataset is a stratified sampled dataset, consisting of speech samples from 5 unique speakers. Two of these were added to include utterances in Sanskrit from speakers with more pronounced influence of their native languages (in Hindi and Tamil). The domain of the training dataset primarily is a speech collection of readings from various well known texts. Further, the speech in the training data is in accordance with the traditional phonetic prescriptions of Sanskrit (Śikṣā). Hence, the remaining three in the out-of-domain test set were added to include utterances from different speech domains, extempore discourse, lecture and radio program, differing from the speech domain in the training set. The radio program is studio produced, while the other two are live recorded.

### 2.1 Challenges in Sanskrit ASR

In this section, we describe various linguistic phenomena that are important to consider when preparing datasets and building ASR systems for Sanskrit with the help of illustrative examples.

---

[3] The URLs of the tools and the list of the texts we use are available in the supplementary material.

**Word Length:** The tokens in Sanskrit texts can be very long owing to "Sandhi" and the lexically productive process of compounding (``Samāsa"). For instance, consider a compound word, वागर्थप्रतिपत्तये(/vāgarthapratipattaye/). It forms a 19 letter word in SLP1 (vAgarTapratipattaye), and is formed by combining the three Sanskrit stems वाक्, अर्थ, प्रतिपत्ति (/vāk, artha, pratipatti/), as per the rules of Sandhi and Samāsa. In Figure 1, we present the distribution of the number of characters (in SLP1 format) per word across the three languages that we experimentally analyse in this work, *viz.*, Sanskrit, Telugu and Gujarati. The plots are normalized with respect to the size of the vocabulary. The average word length is much higher in Sanskrit (10.75) compared to Gujarati (7.79) and Telugu (9.35). Table 2 compares the distribution of number of characters (*wrt* SLP1) per word in the training vocabulary of three ASR datasets. More than 26% of the words in Sanskrit have length exceeding 12 characters.

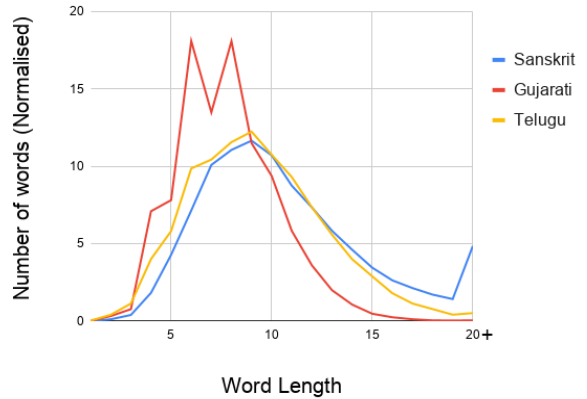

Figure 1: Distribution of Number of characters (in SLP1 representation) per word across Sanskrit, Telugu and Gujarati

| Char count(N) | Sanskrit | Telugu | Gujarati |
|---|---|---|---|
| $N \leq 6$ | 13.79% | 21.27% | 34.11% |
| $6 < N \leq 12$ | 59.56% | 61.60% | 61.82% |
| $N > 12$ | 26.65% | 17.13% | 4.07% |

Table 2: Distribution of number of characters (*wrt* SLP1) per word in three ASR datasets

**Sandhi:** Sandhi can occur between successive words in a sentence or between the lexemes in a compound, or between the morphemes of the

**a) nityaśśabdārthasambandhaḥ**
**b) nityaḥ śabdārthasambandhaḥ**
**c) nityaḥ śabdārtha sambandhaḥ**
**d) nityaśśabdārtha sambandhaḥ**

Figure 2: a) Utterance of a sequence without any pause b) Sandhi split at word boundary c) Sandhi based segmentation d) utterance splits without sandhi

same word. While recognising longer sequences due to sandhi and compounding is a challenge in itself, the external sandhi gives rise to the issue of arbitrary points of segmentation performed by speakers at the time of utterance. Figure 2a shows text-sequence where the sequence contains a word *nityaḥ* and a compound *śabdārthasambandhaḥ* fused together via Sandhi. Further, *śabdārthasambandhaḥ* is a compound with *śabda*, *artha* and *sambandha* as its components. While it is expected to be uttered without any pause after considering *Samhitā* (Aṣṭādhyāyī-1-4-109), a speaker may choose to segment at the lexical boundaries as shown in Figure 2c and Figure 2d. However in doing so, a proficient speaker would prefer a sequence similar to Figure 2a or 2b, rather than Figure 2c or 2d. This is because, the former two, though result in phonetic transformations, preserve the syntactic and semantic validity of the sequence. However, the latter do not preserve the syntactic and semantic validity of the sequence. Similarly, there are cases where there can be phonetic transformations between the bound morphemes and the free morpheme of a word. These transformations do not result in any modification to the word, other than phonetic variations. However, this makes it challenging for an ASR system. The case of Diphthongs is the quite prevalent under these cases.

In Diphthongs, which can occur both at internal or external sandhi, the independent vowel can only occur at the start of a word. Any vowel appearing in the middle of a word either gets converted to a dependent vowel or a diphthong. When a word ending with "ए(/ē/) or ओ(/ō/)" and followed by any vowel, then ending will be changed to either "अय्(/ay/) or अव्(/av/) respectively" or "अ(/a/)". For example विष्णो+ इह(/viṣṇo+iha/) will get converted to विष्णइह(/viṣṇaiha/) or विष्णविह(/viṣṇaviha/).

**Unicode encoding:** The Unicode encoding for the native scripts in Sanskrit, similar to several indian languages, does not preserve the correspondence with the phonemic encoding. Further, mapping the graphemes in Unicode to the corresponding phonemes either suffers from ambiguity and redundancy or often requires multi-grapheme combinations. For instance, consider the word वागर्थाविव(/vāgarthāviva/) in Sanskrit. Here the graphemes in Devanagari 'व ा ग र ् थ ा व ि व' and Roman (v ā g a r t h ā v i v a) do not exhibit a one-to-one mapping with the phonemes. For instance, a single grapheme (e.g., ग) may correspond to 2 phonemes while two graphemes (e.g., र ् in devanagari, 't h' in roman) may correspond to a single phoneme.

## 3 Unit Selection Alternatives for AM and LM

In this section, we discuss different alternatives for identifying units of the language model and the acoustic model that we subsequently employ in our experimental evaluation and analysis.

### 3.1 Alternatives for Graphemic Representation

The Unicode standard for native scripts of Sanskrit : Devanagari, Gujarati and Telugu faces challenge for computational language processing due to redundancy in mappings between phonemes and graphemes as previously discussed. So for Sanskrit, we use Sanskrit Library Phonetic encodings (SLP1) designed by Scharf and Hyman (2011). This encoding gives unique one-to-one mapping to the phoneme. However Gujarati possesses extra native characters such as ઍ(/e/), ઑ(/o/) . Telugu also possesses extra characters such as ఎ(/e/), ఒ(/o/), and ఌ(/l̥/). So we extend SLP1 to fit to the character set of Gujarati and Telugu in our experiments.

### 3.2 Alternatives for Sub-word units

One possibility for deriving subword units for the language modeling is to segment words in Sanskrit based on Sandhi rules. However, Sandhi splitting can change some phonemes corresponding to the words in almost all cases. Consider the word रामायेदम् = रामाय+ इदम् (/rāmāyedam/ = /rāmāya/+/idam/), wherein the vowel ए (/e/) is changed into अ+ इ (/a/+/i/) after performing Sandhi-based splitting. This leads to a mismatch between the speech transcript and the speech audio, potentially creating further complications for

ASR.

### 3.2.1 BPE-based Word Segmentation

Byte pair encoding (BPE) is a simple data compression algorithm that iteratively replaces the frequently occurring subword units with a single unused byte (Gage, 1994). This technique was first adopted to model rare words using subword units in neural machine translation (Sennrich et al., 2016). Interestingly, BPE has been explored for learning new vocabulary for poetry to prose conversion in Sanskrit (Krishna et al., 2019). We consider the benefits of using BPE as a subword unit for Sanskrit ASR.

While BPE is a purely data-driven segmentation strategy, we next present a linguistically motivated segmentation approach that might be aligned with finding syllable units for ASR that are more phonetically compliant. We refer to this technique as *vowel segmentation*.

### 3.2.2 Vowel Segmentation

Splitting the tokens based on vowels and adjacent consonants is inspired by the identification of metres in Sanskrit prosody, where the metre of a verse is identified by using syllable segmentation, followed by identification of syllable weights and it's combinations (Melnad et al., 2013). The syllable weight of a syllable can either be laghu (light)(represented by the symbol |) or guru (heavy)(represented by the symbol S). Syllables with short vowels generally form Laghu and those with loing vowels form a Guru. Also, when a short vowel is followed by a conjunct consonant or Anusvāra (nasal sound /ṃ/) or Visarga (voiceless glottal fricative /ḥ/), the short vowel now becomes Guru. E.g., the *Laghu-Guru* mapping of "अन्यानि संयाति(/anyāni saṃyāti/)" is "SS| SS|". In prior work involving Indian languages for TTS, Kishore et al. (2002) proposed various syllabification rules for words. Herein (with a few exceptions), if a vowel is followed by 3 or more consonants, only the first following vowel is grouped with the preceding vowel to form the subword unit.

Our proposed algorithm for vowel segmentation (VS) is outlined in Algorithm 1. We propose segmenting words at vowel boundaries to extract the units for which alignment with speech is learnt within the ASR system. For acoustic models, an effective unit of a word for ASR would arguably be the syllable (Lee et al., 2013). Representing a word in terms of syllables demands the mapping

---

**ALGORITHM 1:** Vowel segmentation algorithm for Indian languages

**Input:** $word$ in Indian language
**Output:** Vowel segments in output
output = "";
**for** *each graphemic unit $c_i$ in word* **do**
    **if** $c_i$ *is V* **then**
        **if** $c_{i+1}$ *is V* **then**
            output += $c_i$ + " ";
        **else if** $c_{i+1}$ *is C and* $c_{i+2}$ *is C* **then**
            output += $c_i$;
        **else if** $c_{i+1}$ *is C and* $c_{i+2}$ *is V* **then**
            output += $c_i$ + " ";
    **else**
        **if** $c_{i+1}$ *is V* **then**
            output += $c_i$;
        **else if** $c_{i+1}$ *is C and* $c_{i+2}$ *is C* **then**
            output += $c_i$;
        **else if** $c_{i+1}$ *is C and* $c_{i+2}$ *is V and*
        $c_{i+2}$ *is first vowel of the word* **then**
            output += $c_i$;
        **else if** $c_{i+1}$ *is C and* $c_{i+2}$ *is V* **then**
            output += $c_i$ + " ";

---

of a word from graphemes to phonemes. To create syllable units, phonemes are then combined together based on the *sonority sequencing principle* (Clements, 1990). Absence of accurate syllabifiers for Indian languages restricts the use of syllables as units for learning alignment. Our approach produces units which can be viewed as a rough approximation to a syllable. A syllable is composed of three parts *viz.,* onset, nucleus and coda, where nucleus has the highest sonority and is always a vowel. In our approach, the onset is always one or zero consonants and the coda is zero or *n-1* consonants if the nucleus is followed by *n* consonants. It is also observed in the pronunciation of conjunct consonant by professional speakers that the beginning part of conjunct consonant gets associated more with the preceding vowel than the following. We consider nasal Anusvāra (ं), Chandrabindu (ঁ or ঁ), and Visarga (ः) to be part of the consonant set. For example, in Sanskrit, the units for a word उद्यान: (udyāna , park) will be 'उद् या न:(/ud yā na /)' and subword units of the Telugu word తల్లితండ్రులు(/tallitaṃḍrulu/) will be 'తల్ లి తండ్ రు లు(/tal li taṃḍ ru lu/)'.

### 3.3 Vocabulary size as a function of graphemic unit

For each choice of graphemic unit (*viz.* native script and SLP1) described in Section 3.1, we study three different units for the acoustic modeling (AM) in ASR, *viz.*, graphemic unit and vowel segmentation for Sanskrit and also phonemic unit across the two other representative Indian languages *viz.*, Gujarati and Telugu. Whereas, for language modeling (LM), we study word, BPE and VS based units. In Figure 3, we report the vocabulary size based on each of these *three* different unit selections and contrast the sizes with that of two extreme hypothetical systems - one that considers the entire word as a single unit for AM and the other that treats the phoneme as a single unit for AM. Note that while phonetic dictionaries are available for Telugu and Gujarati, our dataset for Sanskrit does not have an accompanying phonetic dictionary. We present the variation in vocabulary size as a function of the graphemic unit (native script vs. SLP1). In both Gujarati and Telugu, we point out that the number of SLP1 graphemic units almost coincide with the number of phonemes, while the native script-based graphemes are much larger in number compared to phonemes.

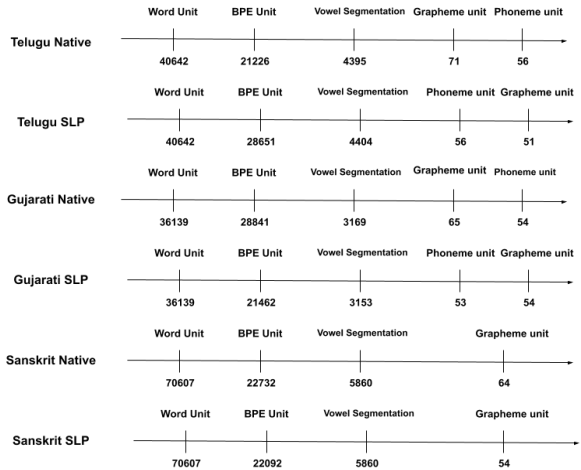

Figure 3: Variation in vocabulary size as a function of the Graphemic unit (native script vs. SLP1) depicting the vocabulary sizes of the whole word units, vowel segment units, BPE units, phonemes and graphemes for ASR data across the three languages including our new Sanskrit ASR data

We can also roughly estimate the extent of data sparsity in terms of vocabulary size in each setting - larger the vocabulary size, higher is the chance of data sparsity. We note that data sparsity is minimal for graphemes and highest for a hypothetical system where whole words are the unit of selection.

## 4 Experiments and Results

**Description of Datasets:** In addition to reporting ASR results on the carefully created वाक्सञ्चयः (/Vāksañcayaḥ/) dataset (described in Section 2), we also contrast through experimental analysis on two other Indian languages, *viz.*, Telugu and Gujarati. For Telugu and Gujarati, we used the publicly available speech corpora released by Microsoft (Srivastava et al., 2018) that contains 36.2/8.7 hours and 33.2/5.8 hours of training/test speech in Telugu and Gujarati, respectively. We use a new train-test split for the Gujarati and Telugu datasets because the original split had overlapping spekaers in their train and test. Our new split ensures that the train-test split have disjoint speakers. Transcript of this corpora was cleaned for orthographic errors. Corpora in these two languages were accompanied by pronunciation lexicons, which we used to build phoneme-based ASR systems to compare against our grapheme-based systems.

**Experimental Setup:** We use the Kaldi toolkit (Povey et al., 2011) for all our ASR experiments. Our acoustic model is implemented using Time Delay Neural Networks (TDNNs) (Peddinti et al., 2015) containing 14 layers. We use 40-dimensional MFCCs as our input features along with 100-dimensional i-vector based speaker embeddings (Saon et al., 2013). We used ngram language models with Kneser-Ney smoothing implemented using the SRILM toolkit (Stolcke, 2002). The language models were trained using both training transcripts from the speech data, as well as additional textual data derived from the Leipzig Corpora Collection for Gujarati and Telugu (Goldhahn et al., 2012) and the Digital Corpus of Sanskrit (Hellwig, 2010) for Sanskrit. The word vocabulary sizes in the lexicons for Sanskrit, Telugu and Gujarati are 76K, 43K and 48K, respectively.

**Results:** Tables 3, 4 and 5, present the WERs from ASR systems built using different choices of AM and LM units using both the graphemic representations (Native and SLP1) for Sanskrit, Gujarati and Telugu, respectively. From Table 3, we see that BPE units[4] and vowel segment units are far superior compared to words as an LM unit for Sanskrit. This is unsurprising given that Sanskrit has a high rates of OOV (44.16%). However, as shown in Tables 4 and 5 the configurations with word based LMs performs the best for Gujarati and Telugu respectively. Gujarati and Telugu have lower OOV rates of 18.63 % and 15.26 %.

| Sr. | Script | LM Unit | AM Unit | WER |
|---|---|---|---|---|
| 1 | Native | Word | Gr. | 40.06 |
| 2 | SLP1 | Word | Gr. | 40.57 |
| 3 | Native | VS | Gr. | 22.55 |
| 4 | SLP1 | VS | Gr. | 22.59 |
| 5 | Native | BPE | Gr. | 21.99 |
| 6 | SLP1 | BPE | Gr. | **21.94** |
| 7 | Native | Word | Gr.+VS | 42.05 |
| 8 | SLP1 | Word | Gr.+VS | 41.36 |
| 9 | Native | VS | Gr.+VS | 24.04 |
| 10 | SLP1 | VS | Gr.+VS | 24.98 |
| 11 | Native | BPE | Gr.+VS | **23.58** |
| 12 | SLP1 | BPE | Gr.+VS | 24.15 |

Table 3: WERs on Sanskrit test set. (Native=Devanagari, VS=Vowel-Split, Gr.=Grapheme)

| Script | LM Unit | AM Unit | WER |
|---|---|---|---|
| Native | Word | Phn. | 18.63 |
| Native | Word | Gr. | 19.17 |
| Native | BPE | Gr. | 24.49 |
| SLP1 | Word | Phn. | **18.26** |
| SLP1 | Word | Gr. | 18.27 |
| SLP1 | Word | Gr. + VS | 19.81 |
| SLP1 | BPE | Gr. | 23.97 |
| SLP1 | BPE | Gr. + VS | 25.08 |
| SLP1 | VS | Gr. | 26.33 |
| SLP1 | VS | Gr. + VS | 26.18 |

Table 4: WERs on Gujarati test set. (Gr.=Grapheme , Phn.= Phoneme)

| Script | LM Unit | AM Unit | WER |
|---|---|---|---|
| Native | Word | Phn. | 21.12 |
| Native | Word | Gr. | 21.25 |
| Native | BPE | Gr. | 26.68 |
| SLP1 | Word | Gr. | **20.75** |
| SLP1 | Word | Phn. | 20.92 |
| SLP1 | Word | Gr. + VS | 22.13 |
| SLP1 | BPE | Gr. | 25.07 |
| SLP1 | BPE | Gr. + VS | 26.3 |
| SLP1 | VS | Gr. | 33.68 |
| SLP1 | VS | Gr. + VS | 36.57 |

Table 5: WERs on Telugu test set. (Gr.=Grapheme , Phn.= Phoneme)

Table 6 shows the distribution of words with 1-4 continuous consonants in all three languages. For Telugu, even though the number of conjunct consonants with $N = 2$ is higher than in Sanskrit, we found on inspecting the audio data that such conjunct consonants are often not enunciated clearly. For example స్వాతంత్ర్యమ్(/svāta**ntry**am/) is pronounced as స్వాతంత్రమ్(/svāta**ntr**am/). However, in Sanskrit, conjunct consonants having 5 consonants together such as कात्स्‍र्न्यम् (/kā**rtsny**am/) are enunciated very clearly and all consonants appear articulated in its pronunciation.

Due to the morphological richness (Kulkarni et al., 2015), inflections and compounds, Sanskrit always has the highest number of rare words. In the training dataset used in the Sanskrit ASR experiments with the vocab size of 70.5K, more than 87.25% words have a frequency less than 3, where as in Telugu and Gujarati training dataset, this is 76.76% and 77.26%, respectively. Clear articulation of conjunct consonants and higher rare word rates makes the BPE and VS based models performs better in Sanskrit than other two languages along with the impact of OOVs.

We observe that use of SLP1 as a graphemic representation schemes performs best for all the three languages. SLP1 is designed to capture the phonemic-graphemic correspondences present in Indic languages. We also find that ASR performance using phonemes is comparable to graphemes for Gujarati and Telugu. In Sanskrit, we observe that purely grapheme-based acous-tic models outperform grapheme+vowel segment-based acoustic models. With the consistent mapping between graphemes and phonemes and the absence of schwa deletion, it is intuitive that grapheme-based models would be most appropriate for Sanskrit. Even though for Sanskrit in some cases Devanagari as a graphemic representation outperforms the SLP1 (Sr. 1,3,9,11 in Table 3), the model that uses SLP1 script always outperforms the other in terms of character error rate.

| N | Sanskrit | Telugu | Gujarati |
|---|---|---|---|
| 1 | 77.30% | 75.77% | 89.37% |
| 2 | 21.41% | 23.27% | 10.06% |
| 3 | 1.26% | 0.96% | 0.56% |
| 4 | 0.03% | 0% | 0% |

Table 6: Number of continuous consonants (N) distribution in three ASR datasets

---
[4]Vocabulary size of 32K is used for BPE which is closest to the vowel split (29,147 entries) for Sanskrit. Performance for varying number of subword units for BPE is presented in Appendix B.

In Sanskrit the pause given between the subwords of a compound word and in between two words varies depending on the fluency of the speaker and the complexity of the text, which can deteriorate the WER. The utterance for 'महान् प्राकारः' /mahān prākāraḥ/ may get recognised as 'महान्प्राकारः' /mahānprākāraḥ/, where two correctly recognised words will be evaluated as one *deletion* and one *substituion* by the evaluation model. Similarly if the audio of 'शोभमानमासीत्' /śobhamānamāsīt/ gets recognised as 'शोभमानम् आसीत्' /śobhamānam āsīt/, then it will be considered as one *insertion* followed by one *substituion*. After negating these two particular errors, we will get 17.79% as the *modulo substitution deletion* WER for our best model of Sanskrit (Sr. 6 of Table 3). The character error rate 3.10% for the best model in Sanskrit also ensures the performance of the model and the quality of the dataset, where as the CER for the best model of Gujarati and Telugu are 5.49% and 5.60% respectively, much higher than Sanskrit.

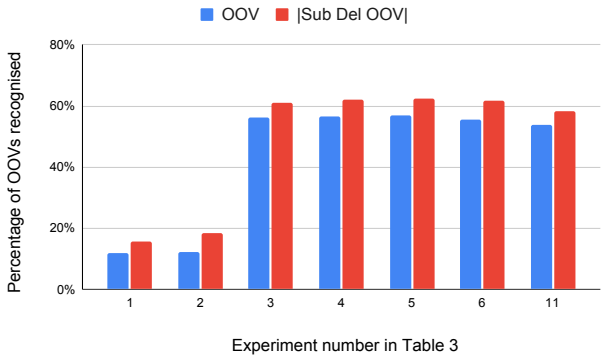

Figure 4: Percentage of OOVs recognised for Sanskrit by the ASR system for the experiments listed in Table 3

**OOV Analysis:** The OOVs at word level are 44.16%, 15.26% and 18.63% for Sanskrit, Telugu and Gujarati, respectively. Figure 4 depicts the percentage of OOVs recovered by different ASR experimental setup described in Table 3, also giving *modulo substitution deletion* OOVs. As we move from word based LM Unit to BPE and VS, system is able to recognize more than 57% of OOVs. For Telugu and Gujarati OOVs recognition rate is 53.67% and 48.61% respectively for their best performing models.

| #  | WER    | Speech Description     |
|----|--------|------------------------|
| 1* | 37.79% | Tamil influenced accents |
| 2* | 37.60% | Hindi influenced accents |
| 3* | 46.27% | Radio Program          |
| 4  | 46.62% | Extempore Discourse    |
| 5  | 51.52% | Live Lecture           |

Table 7: Test results for the Out-of-domain Dataset (* readings from books and transcripts)

**Out-of-domain test set.** Table 7 presents results on the out-of-domain test set described in Section 2. It shows the WERs we can expect from our models when the speakers and content largely vary in domain from our dataset. This test set was sampled for specific speakers and content that qualify as being out-of-domain. These test utterances were evaluated using our best performing Sanskrit ASR models. Speakers #1 and #2 were included, as their utterances show more pronounced influence of their native languages, Tamil and Hindi respectively. It is observed that speaker #1, does not often attempt to distinguish between the pronunciation of the phoneme pairs such as /ta/ and /da/, /ka/ and /ga/, *etc*. This is in congruence with the orthography followed in Tamil, the speaker's native language. Speaker #2's reading was influenced by schwa deletion, *i.e.*, the phenomena of deleting vowel markers accompanying consonants at certain contexts (elaborated in the supplementary material) which is dominant in Hindi. For example, गतवान् /gatavān/ is pronounced as /gatvān/ by this speaker and the ASR system correctly predicts it as /gatvān/. Here, the acoustic model clearly dominates the language model. Speaker #3's utterances are the Sanskrit translation of the Indian government's public outreach program known as 'मन की बात' /man kī bāt/. This widely differs from the training speech dataset and vocabulary of the LM that is used. Similar observation can be made for speakers #4 and #5, albeit for different reasons. Speaker #4 in the discourse tends to use rare words, especially domain-specific proper nouns and derivational verbs, both of which are scarce in our LM vocabulary. Speaker #5 tends to deviate from conventional speech patterns, by providing emphasis on specific words, for the purpose of pedagogy.

## 5 Conclusion

We presented a new Sanskrit speech corpus वाक्सञ्चयः (/Vāksañcayaḥ/) along with a new large-vocabulary ASR system. We explored different unit selection alternatives for both AM and LM, along with a new segmentation approach. We observe that SLP1, when used as the script instead of the native scripts generally results in better performances for Gujarati and Telugu. For Sanskrit, SLP1 based model results in better character error rate and BPE based model with SLP1 giving the best result. Similarly, we observe that vowel split based segmentation consistently yields better performance than word based model and close to results of the best model. Our current dataset and the model are specifically designed to handle data in prose. The inclusion of poetry data would require substantial changes to the system, which we plan to address in the near future. For instance, the poetry data would greatly benefit from insights from Sanskrit prosody. More importantly, the degree of free word orderness in prose and poetry greatly varies in Sanskrit, so much so that an n-gram LM will not be effective (Krishna et al., 2018).

## Acknowledgments

We would like to thank Prof. K. Ramasubramanian, IIT Bombay, for supporting the creation of Sanskrit speech corpus. We express our gratitude to the volunteers who have participated in recording readings of classical Sanskrit texts and helping make this resource available for the purpose of research.

# A Differences between Sanskrit and other Indic languages for ASR

Many Indian languages are known to be derived from Sanskrit (Kulkarni et al., 2010b) and their scripts derived from the Brahmi script (Salomon, 1996; Sproat, 2003), which leads to grapheme-based similarites amongst them. In Figure 5, we illustrate through an example, the spectrum of mapping the native character/grapheme (units) in words across languages; at one end of the spectrum is राम(/rām/) in Hindi mapped to రామ(/rāma/) in Telugu as an example where direct correspondence with the native character exists. Going further in the spectrum are examples for which direct character correspondence does not exist. सीता(/sītā/) in Hindi going to ಸೀತೆ(/sīte/) in Kannada is an instance where there is a change in the ending vowel.

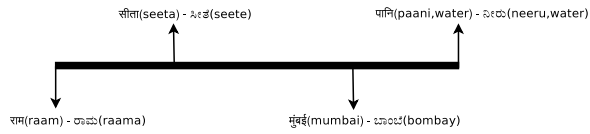

Figure 5: Spectrum of mapping native character/grapheme (units) in words across Indian languages

**Schwa Deletion** The schwa deletion phenomenon plays a crucial role in the north Indian languages. Every consonant by itself includes a short /a/ vowel sound (referred to as "schwa") unless otherwise specified. For example, the letter 'त' in Hindi is pronounced as /ta/. This sound can be associated with any other vowel sound by the use of "Mātras". Mātras are dependent forms of vowels. Schwa is the default vowel for a consonant and hence does not require any explicit Mātra to represent it. Schwa deletion is a phenomenon where implicit schwas of a word are deleted during pronunciation. For example, in Hindi, the proper noun, 'अर्जुन (/arjun/, the name of a person) has schwa deletion after the consonant 'न' and is pronounced as Arjun. This phenomenon is not observed in the South Indian languages. For instance, in Kannada it is pronounced as 'Arjuna'. There is no implicit schwa deletion in Sanskrit as well as in the traditional use of South Indian languages such as Kannada. North Indian languages observe schwa deletion not only at the end of the word, but also in the middle of a word in some cases. For example, the word 'गलती' (/galtī/ meaning mistake) in Hindi observes implicit schwa deletion after the consonant 'ल' (/la/).

ASR becomes challenging because of this phenomenon since the occurrence of schwa deletion is not always explicitly specified in the orthography. For example, the name रामबाबु (/rāmbābu/) has two basic words concatenated to form a name. In Hindi, this name has an implicit schwa deleted at म (consonant sounding 'ma') of राम (/rām/). While constructing phonetic representations for ASR, such deletions introduce ambiguities in pronunciation which could be alleviated by enforcing more consistency between graphemes and phonemes. This same word रामबाबु written in Telugu would be phonetically represented as రామ్బాబు (/rāmbābu/) instead of రామబాబు (/rāmabābu/) which is intuitive. Note that in the former case, there is an addition of '్'(*halant*: an explicit schwa deletion marker) at మ(/ma/). This forces the consonants మ(/ma/) and బ(/ba/) to combine and form a conjunct. In the latter case there is a grapheme consistency across both Hindi and Telugu languages but there is a variation in their pronunciation due to the schwa deletion phenomenon. In contrast, in the case of Sanskrit, since pronunciation is strictly governed by the शिक्षा(/śikṣā/) (Manomohan and Pāṇini, 1938), a treatise on phonetics, schwa deletion is not observed.

# B BPE Experiment Details

In Sanskrit a noun can have 24 to 92 inflections (depending on base word's gender and alternate forms) and a verb can have 90 to 180 inflections. Derivative nouns (*Taddhitas*) and verbs (passive (*Karmaṇi*), *san, ṇic, yaṅ,* etc) are also used often in Sanskrit literature. Due to this morphological richness and frequently occurring compound words, vocab size can be reduced by properly selecting repetitive stems and suffices using BPE by specifying the number of merge operations. Therefore we experimented varying number of subword unit with vocabulary sizes of 2K, 4K, 8K, 16K, 32K and 64K (K=1000). Table 8 shows the varying BPE configuration on our best configuration, i.e graphemes as AM unit and BPE as LM unit. However, the performance of these configurations are comparable irrespective of their BPE vocabulary size.

BPE with vocabulary size of 32,000 stands closes to that of VS, with a vocabulary size of 29,147. Even in this configuration, BPE outperforms VS, as BPE reports a WER of 21.94 as against 22.58 (Table 3 serial no. 4) of VS. However, the BPE configuration with a vocabulary size of 4K reports the lowest WER, which is 0.15% points lower in comparison to BPE with 32K vocabulary size. But the CER for 32K vocabulary size is 3.10% which is outperforming the BPE with vocabulary size of 4K which has a CER of 3.29%.

| Sr. | Vocabulary Sizes | WER |
|---|---|---|
| 1 | 32K | 21.94 |
| 2 | 16K | 21.96 |
| 3 | 8K | 21.83 |
| 4 | 4K | **21.79** |
| 5 | 2K | 22.41 |

Table 8: WERs for Sanskrit on different vocabulary sizes with Graphemes as AM unit and BPE as LM unit using script SLP1 (on the best configuration model).

## C  List of works used in the speech corpus

- Mallinātha's commentary on KumāraSambhavam
- Mallinātha's commentary on Raghuvamśam
- Ādiśaṅkara's Bhaṣyam on Kaṭhopaniṣat
- Ādiśaṅkara's Bhaṣyam on Bhagavadgītā (Chapters 1-9)
- Ādiśaṅkara's Bhaṣyam on Brahmasūtram
- Yogasūtram Vyāsabhāṣya-sahitam
- Ṛṇvimuktiḥ by SaṃskṛtaBhāratī
- Āñjaneya-Rāmāyaṇam by SaṃskṛtaBhāratī
- Kathālaharī by SaṃskṛtaBhāratī
- Bālamodinī stories from SambhāṣaṇaSandeśa by SaṃskṛtaBhāratī
- Samarthaḥ Svāmī Rāmadāsaḥ by SaṃskṛtaBhāratī
- Yugāvatāraḥ by SaṃskṛtaBhāratī
- Prāstāvikam of Swāmī Aḍgaḍānanda's commentary on Bhagavadgītā
- ViśuddhaVedāntaSāraḥ by SaccidānendraSarasvatī
- Vyākaraṇa-Mahābhāṣyam of Patañjali
- Man-Kī-Bāt Sanskrit translation
- Lecture on Lilāvatī
- Extempore Discourse

### C.1  Sources of Recorded Audios

- http://vedabhoomi.org
- https://archive.org/details/Anjaneya-rAmAyaNam
- https://archive.org/details/geethasb
- https://archive.org/details/bAlamodinI-01
- https://archive.org/details/kathA-laharI
- https://archive.org/details/Gita_Shankara_Bhashya-Sanskrit
- https://www.youtube.com/watch?v=LJGjfHHHBoQ
- https://sanskritdocuments.org/sites/manogatam/
- https://archive.org/details/YatharthGeetaSanskritAudio
- https://surasa.net/music/samskrta-vani/#stories_stories_songs

### C.2  Sources of Tools used for Recording, Cleaning and Transcribing the Audios

- ASR Voice Recorder https://play.google.com/store/apps/details?id=com.nll.asr
- Audacity https://www.audacityteam.org/
- oTranscribe https://otranscribe.com/

## D  Computing Infrastructure

- GPU Model Name : GeForce GTX 1080 Ti
- GPU RAM : 12 GB
- CPU Model Name : CPU Intel(R) Xeon(R) Gold 5120 CPU
- Processor Speed : 2.20GHz
- System Memory : 256 GB
- CPU Cores : 56